# Application of Feed-back Connection Artificial Neural Network to Seismic Data Filtering

Noureddine Djarfour[1], Tahar Aïfa[2,*], Kamel Baddari[1], Abdelhafid Mihobi[1], Jalal Ferahtia[1]

[1] Laboratoire de physique de la Terre (LABOPHYT), Université M'hamed Bougara, 35000 Boumerdès, Algeria
[2] Géosciences-Rennes, CNRS UMR6118, Université de Rennes 1, Campus de Beaulieu, 35042 Rennes cedex (France)

[*]Corresponding author: tahar.aifa@univ-rennes1.fr

**Abstract:** The Elman artificial neural network (ANN) (feedback connection) was used for seismic data filtering. The recurrent connection which characterizes this network offers the advantage of storing values from the previous time step to be used in the current time step. The proposed structure has the advantage of training simplicity by back-propagation algorithm (steepest descent). Several trials were addressed on synthetic (with 10% and 50% of random and Gaussian noise) and real seismic data using respectively 10 to 30 neurons and a minimum of 60 neurons in the hidden layer. Both iteration number up to 4000 and arrest criteria were used to obtain satisfactory performances. Application of such networks on real data show that the filtered seismic section was efficient. Adequate cross-validation test is done to ensure the performance of network on new data sets.

*Keywords:* Elman ANN, Gaussian and Random noise, Filtering, training, back-propagation, seismic

# Application du réseau de neurones artificiel récurrent au filtrage des données sismiques

**Abstract:** Le réseau de neurones artificiel (RNA) de type Elman (rétro-connexion) a été utilisé pour filtrer des données sismiques. La boucle de rétro-action qui caractérise ce réseau présente l'avantage de stocker la trace de l'événement précédent pour être utilisée dans le traitement courant. Cette récurrence facilite l'apprentissage de la structure neuronale proposée par rétro-propagation (méthode des gradients).
Des essais ont été réalisés sur des données sismiques synthétiques (avec 10% et 50% de bruit aléatoire et Gaussien) et réelles en utilisant respectivement 10 à 30 neurones et un minimum de 60 neurones dans la couche cachée. Le nombre d'itération fixé à 4000 et un critère d'arrêt ont été utilisés pour satisfaire de meilleures performances. L'application de tels réseaux sur des données réelles montre que le filtrage de la section sismique est efficace. Le test de validation croisée est réalisée pour assurer la performance du réseau sur de nouvelles données.

*Mots clés:* RNA d'Elman, bruit gaussien et aléatoire, filtrage, apprentissage, rétro-propagation, sismique

**Version française abrégée**

**1. Introduction**

Dans la séquence de traitement des données sismiques, la reconnaissance et l'atténuation du bruit est une étape importante. Les méthodes conventionnelles utilisent les propriétés physiques qui différencient le signal du bruit telles que vitesse apparente, nombre d'onde, temps de trajet… par l'application d'algorithmes de filtrage ou de prédiction du bruit par modélisation [1,8,9,13]. Le bruit prédit sera ainsi soustrait des données sismiques. Cependant, de telles méthodes correspondent à une solution dans laquelle le signal désiré est de forme connue. Par conséquent, leurs applications supposent qu'on dispose simultanément de données *a priori* sur le signal et le bruit ce qui n'est toujours pas le cas. Dans la séquence de traitement des données sismiques, la reconnaissance et l'élimination du bruit par les différentes techniques de filtrage restent un défi majeur. Malgré les progrès significatifs effectués au cours de ces dernières années, les performances des filtres proposés sont encore loin d'éliminer totalement le bruit, notamment dans les conditions d'acquisition des données sur des structures géologiques complexes.

Dans ce travail, nous proposons d'utiliser le réseau de neurones artificiel (RNA) pour filtrer les données sismiques. L'utilisation du RNA est motivé par sa capacité à extraire des informations utiles à partir de données hétérogènes ou imprécises. De plus, le modèle RNA permet d'approximer des relations arbitraires complexes non-linéaires et d'obtenir une fonction de transfert. L'accent sera mis sur l'aptitude du RNA à l'apprentissage et à la généralisation sur des données synthétiques et réelles de sismiques réflexions.

**2. RNA et ses applications**

Un réseau de neurones articifiel (RNA) est constitué de neurones élémentaires (''agents'') connectés entre eux par l'intermédiaire de poids jouant le rôle de synapses [11]. L'information est portée par la valeur de ces poids, tandis que la structure du réseau de neurones ne sert qu'à traiter cette information et à l'acheminer vers la sortie (Figure 1a). Le concept du neurone est fondé sur la sommation pondérée du potentiel d'action qui lui parvient des neurones voisins ou du milieu extérieur (Figure 1b). Le neurone s'active suivant la valeur de cette sommation pondérée. Si cette somme dépasse un seuil, le neurone est activé, puis transmet une réponse (sous forme de potentiel d'action) dont la valeur est celle de son activation. Si le neurone n'est pas activé, il ne transmet rien [3,4]. Un RNA est généralement constitué de trois couches. La première couche représente l'entrée; la seconde dite couche cachée, constitue le cœur du réseau de neurones, la troisième représente la couche de sortie (Figure 1a). Un réseau de neurones fonctionne en deux temps : 1- il détermine d'abord ses paramètres suivant un algorithme d'apprentissage (conception) qui fait appel généralement à la rétro-propagation du gradient conjugué [10,12], 2- puis il est appliqué comme fonction de transfert.

**3. Applications**

Nous avons d'abord essayé d'entraîner un réseau de neurone de type « connection-directe » (feed-forward) sans succès par manque de convergence de la rétro-propagation. Calderión-Macías et al. (2000) ont mentionné que le RNA de type "feed-forward" utilisant la rétro-propagation seule ne peut converger, ils ont adopté un apprentissage hybride par la technique du recuit simulé pour entraîner ce genre de réseau. Dans ce travail, nous avons choisi un RNA de type Elman (Figure 1b)(connection récurrente) pour le filtrage des données sismiques. Ce type de RNA possède une couche spéciale appelée «couche de contexte» [5] qui copie l'activité des neurones de la couche cachée, les sorties des neurones de la «couche de contexte» sont utilisées en entrée de la couche cachée. Il y a alors conservation de la trace des activités internes du réseau sur un pas de temps. L'intérêt de ces modèles réside dans leurs capacités d'apprendre des relations complexes à partir des données numériques [11]. Les fonctions d'activation de la couche cachée de la structure proposée sont de type sigmoïde, permettant au RNA de se comporter comme un réseau non-linéaire dans la transformation qu'il réalise. L'algorithme de rétro-propagation du gradient conjugué est utilisé pour entraîner la structure neuronale proposée [2,7].

*Données synthétiques*

Etant donné que l'objectif de cette phase de simulation est le filtrage des données sismiques des bruits aléatoires et gaussiens par RNA, un modèle de subsurface de six couches à stratification horizontale a été utilisé (Tableau 1). Le sismogramme utilisé pour l'apprentissage du RNA ne simule que l'onde P, calculée à incidence normale. A base de ce modèle une section sismique a été calculée par convolution avec une impulsion de Ricker à fréquence centrale 20 Hz, les trajets sont calculés suivant le principe de Snell-Descartes [6]. La section sismique utilisée comme sortie désirée du RNA dans la phase d'apprentissage est donnée en Figure 2a. Remarquons que les simulations ont été réalisées sous Matlab qui n'accepte pas d'indice négatif ou nul pour les distributions de vitesses, engendrant ainsi un décalage temporel pour la première couche qui démarre à 0 ms. Les Figures 2b et 2d qui représentent l'entrée du RNA sont obtenues respectivement par l'addition de 50% de bruit aléatoire et gaussien à la section sismique précédente (Figure 2a). Nous avons réalisé des tests en présence de 10% de bruit aléatoire et gaussien (Tableaux 2a,b), mais nous ne présentons ici que les résultats venant de 50% de bruit. Le nombre d'itération a été fixé à 4000 avec un critère d'arrêt (Tableau 2a). La section sismique filtrée par RNA après apprentissage respectivement pour le bruit aléatoire (Figure 2c) et gaussien (Figure 2e) est compatible avec le modèle sortie désirée (Figure 2a). A part quelques traces dans lesquelles le bruit gaussien (Figure 2e) a été réduit, le filtrage par RNA est plus efficace dans le cas d'un bruit aléatoire. Dans un exemple de performance obtenue (Figure 2f) pendant la phase d'apprentissage avec 10 neurones (Tableaux 2a,b) dans la couche cachée, l'erreur correspond à l'erreur cumulée calculée par comparaison de la sortie RNA avec la sortie désirée. La performance avec 4000 itérations prédit une erreur croissante quand le nombre de neurones augmentent. On remarque que la convergence est plus rapide lorsqu'on a affaire à du bruit aléatoire (Figure 2f). Si l'on compare les figures 2c,e, la qualité du filtrage est meilleure dans le cas du bruit aléatoire.

*Données réelles*

Deux types de données ont été préparées par filtrage dans le domaine spectral, sous forme de couples constitués d'entrée (traces d'un point de tir) et de sortie désirée (traces filtrées). L'un pour réaliser l'apprentissage et l'autre pour tester le réseau obtenu et déterminer ses performances. Les données d'apprentissage (Figure 3a), composées d'une collection de 48 traces à 2000 échantillons par trace, traitées en amplitudes égalisées, contiennent plusieurs types de bruits (bruits aléatoires, arrivées directes, réfractées, ground roll,…) qui ont été filtrés dans le domaine (f-x) (Figure 3b) pour fournir au RNA un exemple représentatif de bruits qu'il doit apprendre à identifier et à atténuer.

Plusieurs essais d'apprentissage ont été réalisés (Tableau 2c). Les performances obtenues montrent la simplicité d'apprentissage de ce type de RNA par l'algorithme de rétro-propagation du gradient conjugué et sont donc satisfaisants (cf Figure 3c).

Deux tests ont ensuite été réalisés. Le premier qui est un test d'apprentissage, a été accompli en utilisant les données d'apprentissage (Figure 3d) pour constituer la sortie du réseau après apprentissage, en lui présentant les traces bruitées utilisées dans cette phase. En comparant la sortie du réseau après apprentissage (Figure 3d) avec les données bruitées (Figure 3a), on remarque la reproductibilité de la sortie désirée (Figure 3b) par l'atténuation du bruit. Le second est un test de généralisation. Dans ce cas, le deuxième modèle de données (Figure 4a) constitue l'entrée du RNA entraîné, tandis que les données filtrées dans le domaine (f-x) (Figure 4b) sont la sortie désirée que l'on comparera à la sortie du RNA. Nous avons obtenu une section sismique (Figure 4c), où l'on observe l'atténuation du bruit si on la compare à la section filtrée donnée en figure 4b qui est le résultat attendu. Des horizons peuvent être isolés dans la partie supérieure de la section (<1200ms) où le filtrage était effectif : réduction des zones énergétiques (<400ms), atténuation du ground roll, des arrivées directes et réfractées à différents niveaux.

**4. Conclusion**

L'apprentissage du RNA constitue un moyen de synthétiser automatiquement une fonction de transfert non-linéaire entre les données bruitées et les données filtrées qui permet au RNA de se comporter comme un filtre. Le RNA de type Elman offre l'avantage de simplicité d'apprentissage par la rétro-propagation du gradient conjugué.
La parfaite concordance entre la sortie du RNA en phase de généralisation et la section filtrée, résultat attendu, témoigne de la réussite de la structure neuronale proposée en phase de généralisation. Ces résultats obtenus démontrent clairement la faisabilité de la méthode, ainsi que l'intérêt d'exploiter la capacité de filtrage des données sismiques offerte par l'utilisation de l'approche neuronale.

**1. Introduction**

The filtering of noise is one of the crucial steps in a data processing sequence, especially in light of the complexity of observations and the interference of noise types. Noise occurs not only during the initial phase of data acquisition from different sources (e.g. intrumental, natural) and processing related to techniques of sampling and methodology of filtering, but can also be visible in the final document. Methods used to attenuate the noise either by exploiting physical properties which differentiate noise from signal, such as apparent velocity, wavelet number, travel time… by the use of filtering algorithms or by predicting the noise through modelling techniques [8,9,13]. Such predicted noise will be subtracted from the recorded seismic data. However, such methods correspond to a solution when the desired signal is of a known shape. Consequently, their application simultaneously assumes *a priori* information on signal and noise. Such information can be a minimum phase analytical signal (e.g. Ricker) we usually use for seismic for instance.

In spite of significant progress regarding the performance of the proposed filters, we are still far from reducing the noise completely either from synthetic or real data. So far there is no noise attenuation technique that works universally in all acquisition data. Therefore, recognizing and removing the noise remains one of the great challenges in signal processing. The method proposed in this paper is an attempt to filter the noise by means of the artificial neural network (ANN). The use of ANN is motivated by its remarkable ability to extract useful information from heterogeneous or inaccurate data. Moreover, the ANN can approximate arbitrary complicated non-linear relationships and get a good non-linear transfer function. The emphasis is placed on the capacity of the ANN for training and generalization.
We will first address the ANN by means of several tests performed on synthetic and real seismic shot data, then make an adequate cross-validation and compare the quality of filtering at different stages of training and generalization. The filtering by means of ANN has not yet been applied to seismic processing sequence.

## 2. Artificial Neural Network principles and its application

An ANN is an information processing paradigm that is inspired by the way biological nervous systems, such as the brain, process information. The most basic element of the human brain is a specific type of cell, which provides us with the ability to remember, think, and apply previous experiences to our every action [11]. These cells are known as neurons. The power of the brain comes from the number of these basic components and the multiple connections between them. The key element of this paradigm is the novel structure of the information processing system. It is composed of a large number of highly interconnected processing elements (neurons) working in unison to solve specific problems [4]. Each neuron is linked to certain of its neighbours with varying coefficients of connectivity that represent the strengths of these connections [3]. Each neuron transforms all signals that it receives into an output signal, which is communicated to other neurons. For example, the artificial neuron i multiplies each input signal by the synaptic coefficient $w_i$, and adds up all these weighted entries in order to obtain a total simulation. Using an activation function f (or transfer function), it calculates its activity at the output, which is communicated to the following neurons (Figure 1a).

Neurons are grouped into layers. In a multi-layer network there are usually an input layer, one or more hidden layers and an output layer (Figure 1b). The layer that receives the inputs is called the input layer. It typically performs no function on the input signal. The network outputs are generated from the output layer. Any other layers are called hidden layers because they are internal to the network and have no direct contact with the external environment. The 'topology' or structure of a network defines how the neurons in different layers are connected. The choice of the topology to be used depends closely on the properties and the requirements of the application.

*Artificial Neural Network's training*

The most widely used training method is known as back-propagation method. This later produces a least-square fit between the actual network output and desired results by computing a local gradient in terms of the network weights [12]. The design of neural network models capable of training originates in the work of the neurophysiologist Hebb (1949) [11]. Let us start from the simplest and best-known of facts: the neurons are connected to each other by synapses. Hebb made the assumption that the force of a synapse increases when the neurons which it connects act in the same way at the same time. Conversely, it decreases when the neurons have different activities at the same time. Consequently the intensity of this force varies, more or less, according to the simultaneous activity of each inter-connected neuron. A learned form will correspond to a state of the network in which certain neurons will be active while others will not. It is the same principle which is applied to artificial neuron training. All the neurons that are connected to each other constitute a network. When one presents to the network a form to be learned, the neurons simultaneously enter into a state of activity which causes a slight modification of the synaptic forces. What follows is a quantitative reconfiguration of the whole of the synapses, some of them become very strong (high value of synaptic force), and the others become weak. The learned pattern is not directly memorized at an accurate location. It corresponds to a particular energy state of the network, a particular configuration of the activity of each neuron, in a very large case of possible configurations. This configuration is supported by the values of the synaptic forces [10,12].

Let $y_j^s$ represent the output of the j[th] neuron at layer s, $w_{ij}^s$ is the weight connecting the i[th] neuron in layer s-1 to the j[th] neuron at layer s and $o_j^s$ is the weighted sum at the input of the j[th] neuron of layer s. This output can be expressed by the following equation [7]:

$$y_j^s = f(o_j^s) = f(\sum_{i=1}^{N} w_{ji}^s \cdot y_i^{s-1} - b_j) \quad (1)$$

Where N is the number of neurons in the layer s-1. This relation allows, by knowing the input of the first layer of the network, to gradually calculate the value of the global output of the network, thus ensuring the forward propagation. When one compares this output with the desired output [7], one can calculate the error function, given by:

$$e = \frac{1}{2} \sum_{k=1}^{M} (y_k - \bar{y}_k)^2 \quad (2)$$

where $y_k$ is the desired output, $\bar{y}_k$ the obtained output of the ANN and M is the number of neurons in the output layer. In the back-propagation method, the direction, in which weights $w_{ij}^s$ are updated, is given by the opposite gradient of e with respect to every element of the weight. The weights update starts in the output layer and processing will be spread towards the first layer, where the root of the term "back-propagation" comes. The back-propagation algorithm consists in minimizing e. In our case, we used the gradient steepest descent [3] to minimize e.

With each example, one has to modify the synaptic weights so that we may drastically reduce the value of e. The modification of the synaptic weights is carried out using the following relation:

$$\Delta w_{ij}^s = -\mu \cdot (e_j^s \cdot y_i^{s-1})_n + (\Delta w_{ji}^s)_{n-1} \quad (3)$$

where μ is known as the learning rate parameter to speed the ANN training and is usually a small number [2], say between 0 and 0.5. The amount $e_j^s$ is the local error of j[th] neuron in the layer s defined as:

$$e_j^s = \bar{f}(o_j^s) \sum_{k=1}^{N} e_k^{s+1} w_{kj}^{s+1} \quad (4)$$

In these equations, $\bar{f}$ is the derivative of $f$,
Weights and threshold terms are first initialized to random values [2]. In general, there are no strict rules to determine the network configuration for optimum training and prediction.

*Seismic data filtering*

One can model the seismic signal by means of the following equation [6,8,9]:
$$T(t) = W(t) * R(t) + B(t) \quad (5)$$

where (*) represents the convolution product, T(t) the actually measured signal, W(t) the emitted signal, R(t) the reflectivity set (the effective signal to be evaluated) and B(t) the seismic noise. Generally, one distinguishes the natural noise, which exists apart from all seismic activities and the artificial noise, produced by the recording or the processing of seismic data. These two types of noise can be coherent or random. All the matter of filtering consists of starting from a measured function T(t), to obtain a filtered function which estimates as adequately as possible the convolution product of W(t)*R(t). Many efforts in signal processing were directed towards the definition of powerful filtering tools, which must be flexible and able to adapt with the variation of the noise type. In fact, very encouraging results are available today for the filtering and the improvement of the signal to noise ratio. These results are obtained either by temporal or frequency filtering. Consequently, several filters are proposed which can be mainly the adapted filter of Wiener, Kalman types, etc. [1,8,9,13]. All the conventional techniques for noise attenuation fundamentally require *a priori* information on characteristics of signal and noise. This information is usually not well known and the parameters which characterize the signal and noise can often only be estimated inaccurately. Here we make an attempt to overcome some of these limitations by designing a network that can learn characteristics from signal and noise during processing in the training phase. In fact, by the use of ANN technique, there is no *a priori* information on signal and noise characteristics.

## 3. Applications

Because a three-layered neural network could approximate any arbitrary complicated non-linear relationship, we tried to train feed-forward ANN unsuccessfully due to non-convergence of back-propagation. Calderión-Macías et al. (2000) mentioned that feed-forward ANN using back-propagation alone could not converge, they used hybrid training by very fast simulated annealing technique to train this kind of network. In the present study we adopted a three-layered feedback architecture of neural network (Figure 1b), which is of Elman network type. It is one hidden layer with feedback connection from the hidden layer output to its input that constitutes the context representation layer. This feedback loop allows Elman networks learning to recognize and generate temporal or spatial patterns. This internal looping ensures a recirculation of information inside the hidden layer of this Elman ANN. In fact, the delay in this connection stores values from the previous time step, which can be used in the current time step [5]. In networks of this type, the hidden layer is activated simultaneously by the input layer and the context layer. The result of processing of this hidden layer at its output is communicated to output of Elman ANN and will be stored in the context layer. By analogy with human brain thinking, this delayed information leads to hesitations before the making of a final decision by a human brain.

The activation functions of the hidden layer are of log-sigmoid type given by the following equation [2,3]:

$$f(x) = \frac{1 - e^{-2\alpha x}}{1 + e^{+2\alpha x}} \qquad (6)$$

where α is a parameter which controls the steepness of the function near x=0.
  This function can produce outputs with a reasonable discriminating power and is a differentiable function which is essential for the back-propagation of errors. The number of neurons in the hidden layer depends on the desired optimum performance, which could be selected on a trial and error basis.
All further simulations and applications on real data were performed under Matlab environment.

*Synthetic data set*
    As an application for the method described above, we designed a synthetic subsurface model that is 500m wide and 500m deep. Since the main goal of this phase of simulation is to train the Elman ANN to filter seismic data from random and Gaussian noise in which subsurface consists of six homogeneous horizontal layers. A linearly increasing velocity structure is given with interval velocities varying with depth from 1200 m.s$^{-1}$ to 4000 m.s$^{-1}$ (Table 1). The synthetic seismograms used for network training include only P-wave primary reflections computed at normal incidence. Using this model, a set of trace gathers was

computed using convolution equation (5), for which ray paths are computed by Snell-Descartes' law. The convolved seismic wavelet corresponds to a Ricker wavelet with a central frequency of 20 Hz. The synthetic seismic section used as the desired output of the ANN in the training phase is shown in Figure 2a. We may notice that Matlab does not agree with index lower or equal to zero for the velocity distributions, generating thus a temporal shift for the first layer beginning with 0 ms. In the input of the ANN (Figure 2b), the previous seismic section (Figure 2a) was mixed with 50% of random noise. We also did tests on synthetic data with additional 10% of random and Gaussian noise (Tables 2a,b), but here we represent and discuss only results coming from 50% of noise. The ANN with different numbers of neurons in the hidden layer was trained using the back-propagation algorithm [2] with an iteration number of 4000 (Table 2a).

In an example of obtained performance, during the training phase with 10 neurons (Table 2a,b) in the hidden layer (Figure 2f), the plotted error corresponds to the accumulated errors obtained from comparing the filtered shot gather (namely the desired output) with the network output. Two criteria to arrest the calculation are adopted: if the performance is around $10^{-6}$ or if the number of iteration is 4000. Different tests were then performed with different numbers of neurons in the hidden layer, the prediction error increases as the number of neurons grows, if we keep 4000 iterations. One can notice that the performance curve for random noise converges more quickly than that for the Gaussian noise (Figure 2f). If we compare both Figures 2c,e , the quality of filtering is better on random noise.

The result of filtering the seismic section by the ANN after training is shown in Figure 2c. The obtained seismic section by ANN after training (Figure 2c) is compatible with the seismic model which is also the desired output (Figure 2a). When we add 50% of Gaussian noise to our model (Figure 2d), one can see that the obtained section using ANN after training (Figure 2e) is similar to the seismic model (Figure 2a). We therefore demonstrated that the ANN technique clearly improved the seismic section, even if some noise with reduced amplitude still remains.

The ANN technique was also applied using the same procedure, with the same model, to filter Gaussian noise at different rates of noise, respectively 10% and 50%. The example, in which 50% of Gaussian noise was added to the seismic section, reveals that in the second (~120ms) and the fourth (~320ms) horizon the signal to noise ratio is weak. According to our model, i.e. for a constant rate of noise (here 50%), when the velocity contrasts are lower, the signal to noise ratio is weaker and conversely. The result of the ANN application (Figure 2e) is in agreement with the model itself (Figure 2a). It shows, apart from a few traces where the noise has been reduced significantly, a similarity with the model which can be distinguished (Figure 2a). Comparing this result with the one obtained with 50% of random noise, we can conclude that better filtering is efficient in presence of random noise.

*Real seismic data*

The three following steps must be given chronologically if we wish to implement the use of ANN technique to filter real seismic data.

*(i)- Data preparation*

In order to implement the ANN to filter seismic data, two sets of models are used. The first model is used to train the proposed network structure, while the second one is used to test whether the network can filter the new data using the derived weights. The seismic data used in this study were extracted from a survey of an oil field in the South of Algeria. Each model, which constitutes a record shot gather using a Vibroseis source and a 48 trace record with a shift of 20m for the source and receivers, was processed with equalized amplitudes. The sampling rate is 2ms and each trace contains arrivals between 100ms and 4000ms. In order to prepare the data for training, each shot was subjected to an FFT, its frequency components were analyzed on the x direction. If the events recorded within the analyzed traces show coherent aligned signals, the horizontal frequency will be regular along the x direction. Deviations from such regular coherency are caused by the presence of noise. A predictive operator is built in the x direction to reduce the effect of noise by horizontalizing the coherent events. Such operation will isolate the noise from the signal, it will be removed easily by subtraction from the original data. Finally the result will be given in the temporal domain through an inverse FFT algorithm. As an example one can see that shot gather (Figure 3a) shows different kinds of noise (random noise, direct and refracted waves, ground roll) which are reduced significantly by application of such a filtering procedure (Figure 3b). The projective filter in the (f-x) domain separates the signal, assuming that it is predictable in x, from non-predictable noise within all the frequency spectra. We note that the ground roll has been severely reduced, direct and refracted waves were also filtered so that the first horizons can be easily identified between 400 and 900ms (Figure 3b). We may also notice that energetic waves were also reduced before 200ms. We distinguish different levels of reflections between 200 and 500ms. Some horizons are also detected, for instance at around 1300 and 1800ms (Figure 3b) even if they were lacking or presenting a very low signal to noise ratio in the original data (Figure 3a).

*(ii)- Training phase*

The shot gather selected from the training set (Figure 3a) together with the filtered one (Figure 3b) are used as the desired output of the ANN in the training phase. Since the efficiency of any network application depends mainly on the database used for training and testing the network, the training data set should contain all the possible noise types that can appear in the field.
The ANN with different numbers of neurons in the hidden layer was trained using the back-propagation method. Weights were randomly initialized and the learning rate is at 0.05 [10,11], the number of neurons in the hidden layer is 50, 60, 70 and 90 and the generalization performance is reported in Table 2c. All networks were trained for an identical maximum number of iterations (4000). The performance obtained during the training phase with 60 neurons in the hidden layer is shown in Figure 3c. Tests performed with different numbers of neurons in the hidden layer show that the prediction error decreases as the number of neurons grows, if we keep 4000 iterations (Table 2c).
However, final training errors are similar if we deal with a number ranging between 60 and 90 neurons in the hidden layer. For 50 neurons the performance is not really accepted.

To test the ability of learning of the proposed ANN structure we conducted tests on its capability to produce outputs for the set of inputs that were used in the training. The result obtained is shown on section of Figure 3d. If we compare the output of the network (Figure 3d) after the training, using the noisy shot gather (Figure 3a), with the filtered one (Figure 3b), one can easily notice that the network behaves like a filter. It significantly reduced several types of noise (random noise, direct and refracted waves, ground roll). The network fits the training data accurately and the 60 neurons in the hidden layer are therefore able to perform the task of training satisfactorily.

*(iii)- Generalization phase*

Once the training is carried out, i.e. the architecture and the activation functions, the synaptic weights of the network are fixed, the obtained network is used like a classical non-linear function. Now that we have tested the network obtained on the training data, it is important to see what it can do with new data, not mentioned before. If this ANN does not give reasonable filtered outputs for this test set, the training period is not achieved. Indeed, this testing is critical to ensure that the network has not simply memorized a given set of data but has learned the task of data filtering.

The test for its generalization ability is carried out by investigating its capability to filter the new data that were not included in the training process. In this study, for a test network generalization, we plotted output sets and desired outputs of data that were chosen for this stage of network development. To test the trained ANN, new data were used with some noisy events (Figure 4a). One can see that the noise is attenuated on the obtained output ANN (Figure 4c). In this case the trained network is said to have a good generalization performance.

Let's compare both ANN output in the generalization phase (Figure 4c) and the obtained filtered section (Figure 4b) which is the desired ANN output, using the previous procedure in the data preparation. Isolated horizons can be seen mostly in the upper part of the section (<1200ms) where filtering was effective: reduction of energetic areas (<400ms), attenuation of direct and refracted waves and ground roll at different levels.

## 4. Conclusion

In this work the neural networks of Elman type were addressed to filter seismic data. The results show that networks could establish relationship between noisy and filtered traces, based on their ability of approximation and adaptation. The training of the network can be considered as a means to synthesize automatically a function (control mapping), generally a non-linear one, which plays a role of filter.

In our case the learning ANN algorithm used a steepest descent technique which is based on straight downhill in the weight space till the first valley is reached. This makes the choice of initial starting point in the multi-dimensional weight space critical. To overcome this limitation, the training process was repeated a number of times with different starting weights. The excellent results indicate that training of this network was done successfully, and the Elman network has the ability to obtain strict convergence for a complicated set of data in case of filtering of seismic data. Moreover, the method is able to reproduce features that were not included in its training set. In fact, based on the results of the ability of generalization, the application of the ANN technique in a same seismic exploration area produced reasonable signal to noise ratio on data sets that were not learned.

Several tests were performed with synthetic data in the presence of two kinds of noise (random and Gaussian) with rates of 10% and 50%. The results show that the used ANN structure (i.e. Elman type) was able to recognize and reduce the noise. The same results were also obtained on real seismic data.

We think that feedback neural networks, with their ability to discover input-output relationships will increasingly be used in engineering applications, especially in petroleum engineering usually associated with intrinsic complexity. The results obtained demonstrate that the neural network can detect several types of noise in data contaminated by multiple events classified as noise (e.g. direct and refracted waves, ground roll, random noise) and reduce them significantly.

The main advantage of using ANN for filtering seismic data is that once the ANN has been trained, it has the ability to quickly filter new data.

**Acknowledgements**

We thank both the two anonymous referees who suggested some remarks allowing to improve the final version of the manuscript.

**Figure captions**
**Liste des figures**

Figure.1. Architecture of an artificial neuron and a multilayered feedback neural network. (a) Artificial neuron (b) Multilayered artificial neural network
Figure 1. Configuration d'un neurone artificiel et un réseau de neurone artificiel à connexion récurrente. (a) neurone artificiel (b) réseau de neurones artificiel multicouches.

Figure 2. Training on synthetic data. (a) used model (see Table 1), considered as desired output. ANN input (same section as in Figure 2a) with 50% of random (b) and Gaussian (d) noise respectively with their ANN output after training (c, e). (f) Mean square error (MSE) vs. number of iterations for a training stage. Full (dashed) line curve corresponds to 50% of added random (Gaussian) noise.
Figure 2. Apprentissage sur données synthétiques. (a) Modèle utilisé (voir tableau 1), considéré comme sortie désirée. Entrée du RNA (même section qu'en figure 2a) avec 50% de bruit aléatoire (b) et gaussien (d) et leur sortie du RNA respectivement après apprentissage (c,e). (f) Erreur quadratique moyenne en fonction de nombre d'itérations, dans la phase d'apprentissage. Courbe pleine (tiretée) correspondant à 50% de bruit aléatoire (gaussien) ajouté.

Figure 3. (a) Full shot gathers. Input of ANN. (b) corresponding filtered shot gather which is the desired output for the ANN in the training phase. (c) The performance obtained during the training phase. (d) Output of ANN in the training test. Rn: random noise, Dw: direct wave, Rw: refracted wave, Gr: ground roll.
Figure 3. (a) Point de tir. Entrée du RNA. (b) point de tir filtré correspondant qui représente la sortie désirée du RNA. (c) Performance obtenue pendant la phase d'apprentissage. (d) Sortie du RNA dans le test d'apprentissage. Rn : bruit aléatoire, Dw : onde directe, Rw : onde réfractée, Gr : onde de surface.

Figure 4. (a): full shot gather. Input of the ANN. (b): corresponding filtered shot gather which is the desired output of ANN during the generalisation test. (c): output of ANN during the generalization test. Same notations as in Figure 3.
Figure 4. (a): point de tir. Entrée du RNA. (b): point de tir filtré qui correspond à la sortie désirée du RNA pour le test de généralisation. (c): sortie du RNA pendant le test de généralisation. Mêmes notations qu'en figure 3.

**Table captions**
**Liste des Tableaux**

Table 1 : Parameters of simulation used to generate the training data for ANN.
Table 1 : Paramètres de simulation utilisés pour générer les données d'apprentissage du RNA.

Table 2 : Results of training on synthetic data with random (a) and Gaussian (b) noise and on real data (c).
Table 2 : Résultats de l'apprentissage sur des données synthétiques avec du bruit aléatoire (a), gaussien (b) et sur des données réelles (c).

| No of layer | Velocity (m.s$^{-1}$) | Thickness (m) |
|---|---|---|
| 1 | 1200 | 30.6 |
| 2 | 1700 | 42.5 |
| 3 | 2100 | 105 |
| 4 | 2700 | 135 |
| 5 | 3200 | 160 |
| 6 | 4000 | 200 |

Table 1

| Percentage of noise | Number of neurons in hidden layer | Number of iterations | Performance |
|---|---|---|---|
| 10% | 10 | 4000 | 1.66204e$^{-06}$ |
|  | 15 | 4000 | 3.84408e$^{-06}$ |
|  | 20 | 4000 | 6.30692e$^{-06}$ |
|  | 25 | 4000 | 9.92197e$^{-06}$ |
|  | 30 | 4000 | 1.39624e$^{-05}$ |
| 50% | 10 | 4000 | 2.36196e$^{-06}$ |
|  | 15 | 4000 | 2.87078e$^{-05}$ |
|  | 20 | 4000 | 4.60794e$^{-05}$ |
|  | 25 | 4000 | 4.98179e$^{-05}$ |
|  | 30 | 4000 | 8.00859e$^{-05}$ |

(a)

| Percentage of noise | Number of neurons in hidden layer | Number of iterations | Performance |
|---|---|---|---|
| 10% | 10 | 4000 | 1.66596e$^{-06}$ |
|  | 15 | 4000 | 3.63946e$^{-06}$ |
|  | 20 | 4000 | 6.33773e$^{-06}$ |
|  | 25 | 4000 | 9.83710e$^{-06}$ |
|  | 30 | 4000 | 1.09455e$^{-05}$ |
| 50% | 10 | 4000 | 4.65452e$^{-06}$ |
|  | 15 | 4000 | 9.39854e$^{-06}$ |
|  | 20 | 4000 | 9.50304e$^{-06}$ |
|  | 25 | 4000 | 1.39683e$^{-05}$ |
|  | 30 | 4000 | 2.83009e$^{-05}$ |

(b)

| number of neurons in hidden layer | Number of iterations | Performance |
|---|---|---|
| 50 | 4000 | 368.4620 |
| 60 | 1201 | 2.33044e$^{-08}$ |
| 70 | 794 | 1.77530e$^{-08}$ |
| 80 | 648 | 6.17601e$^{-08}$ |
| 90 | 573 | 1.09518e$^{-08}$ |

(c)

Table 2

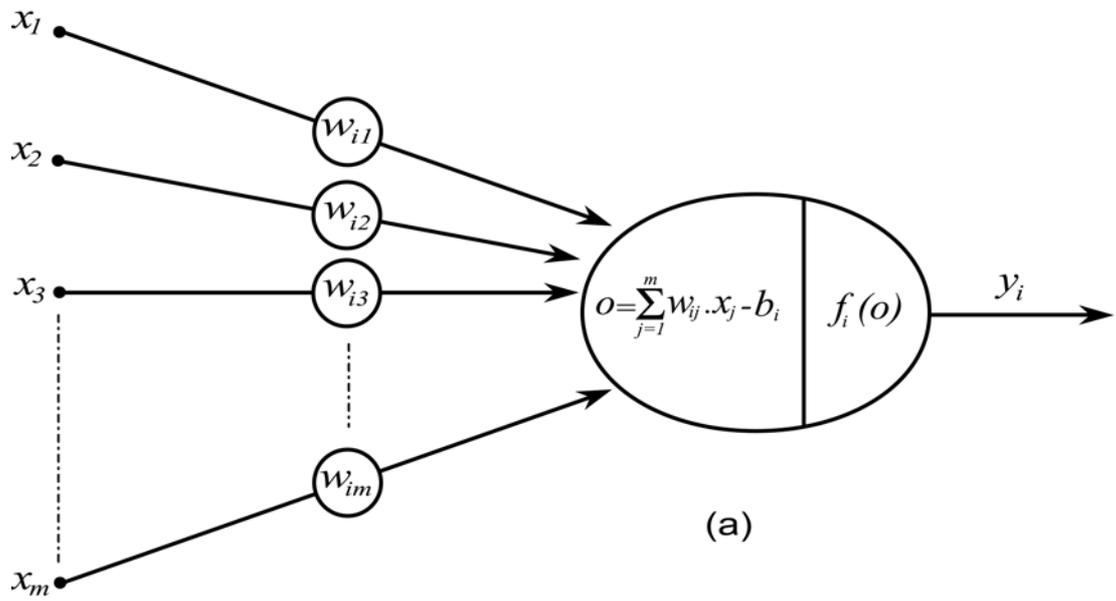

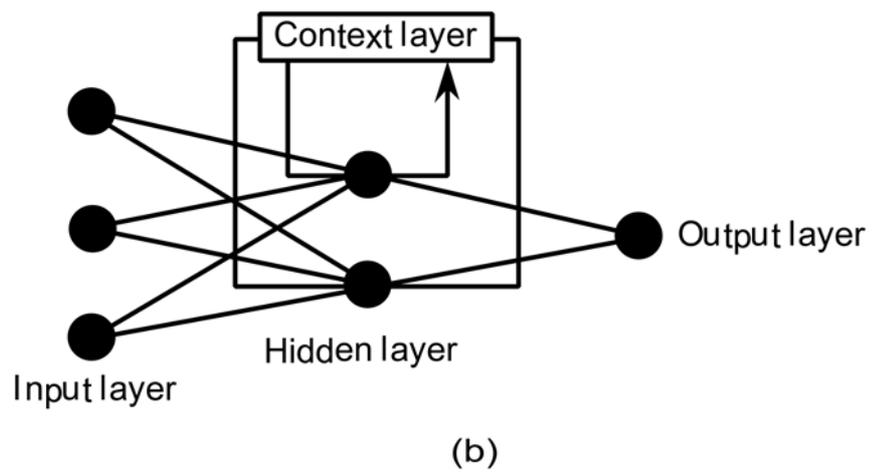

Figure 1

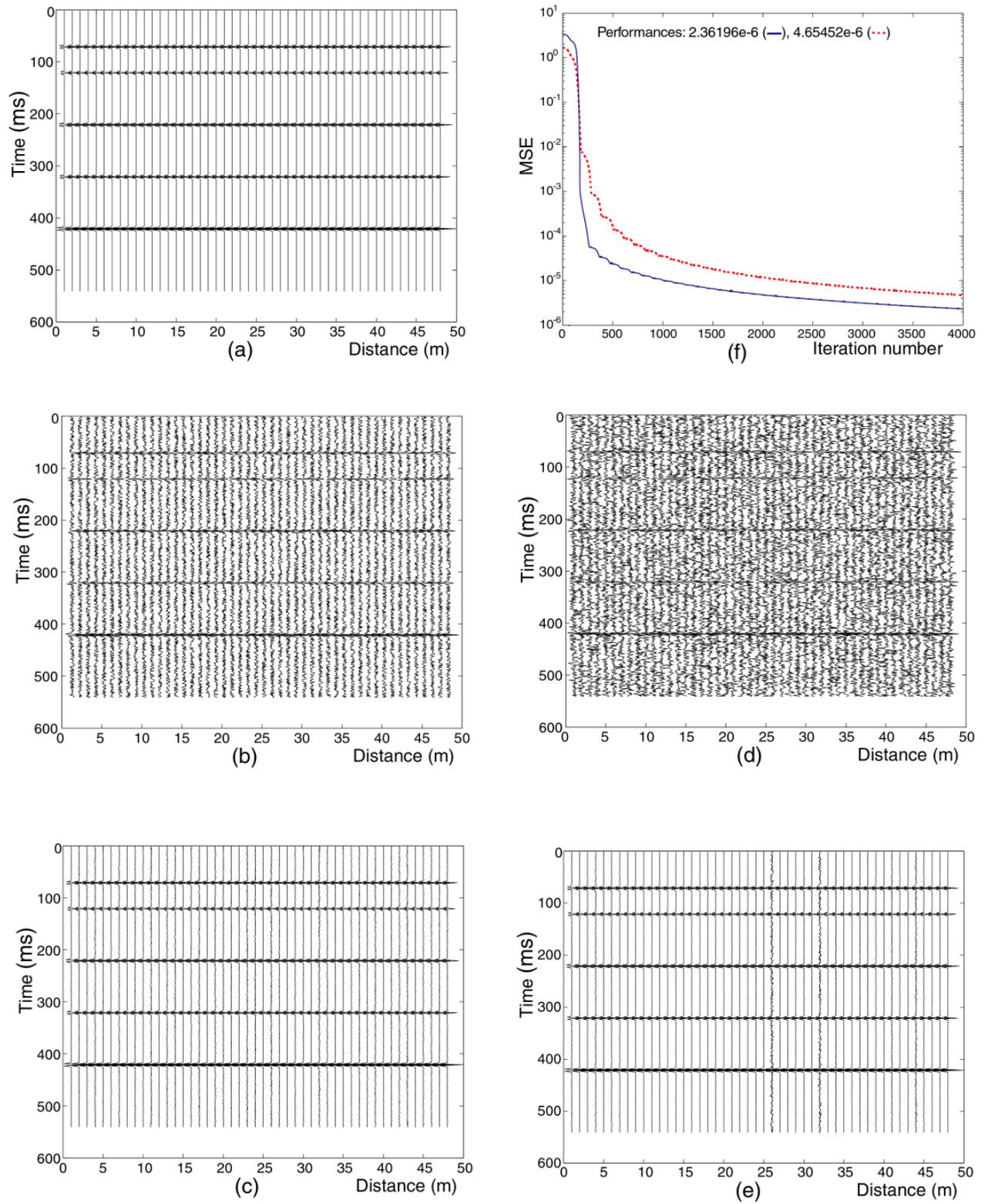

Figure 2

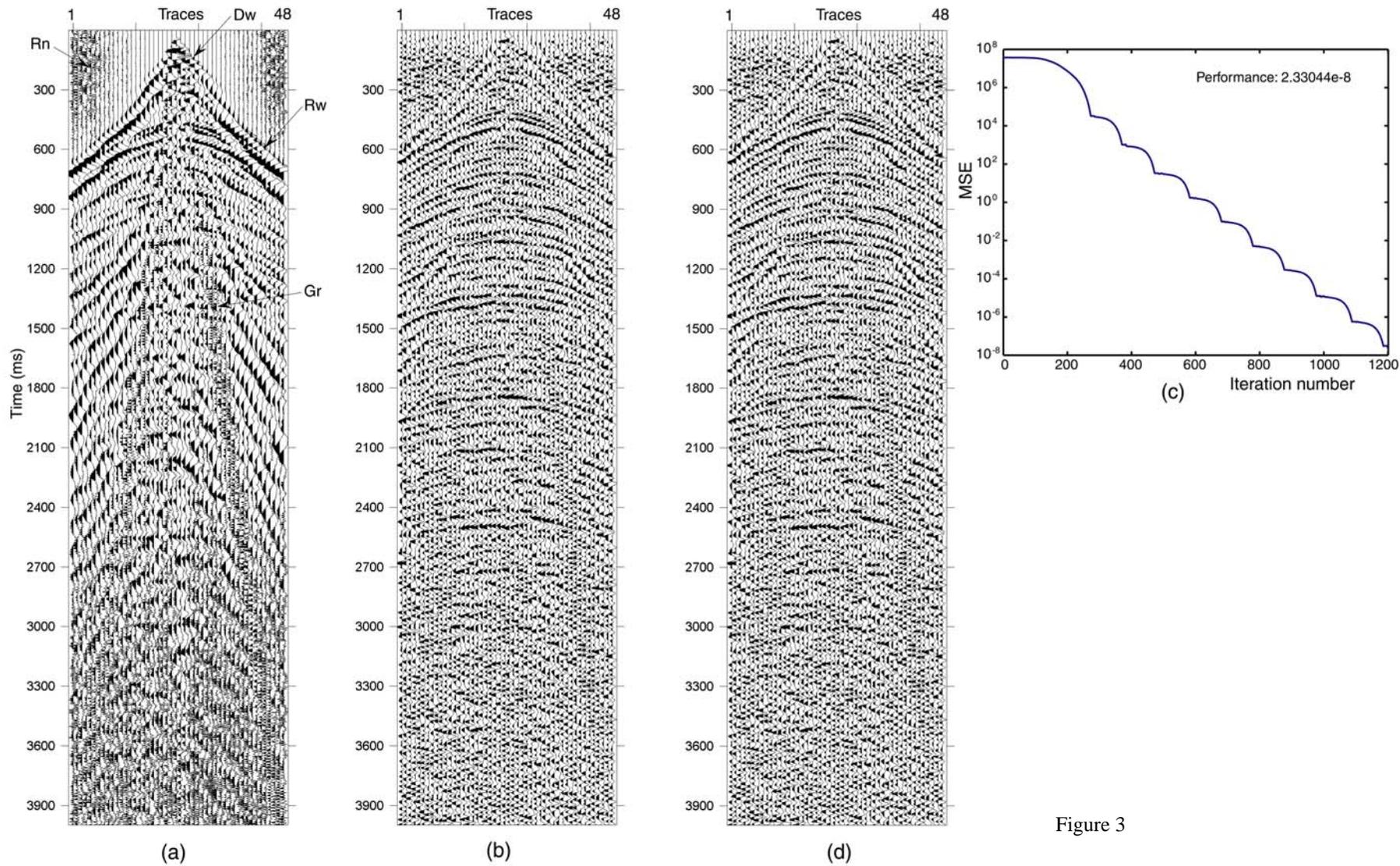

Figure 3

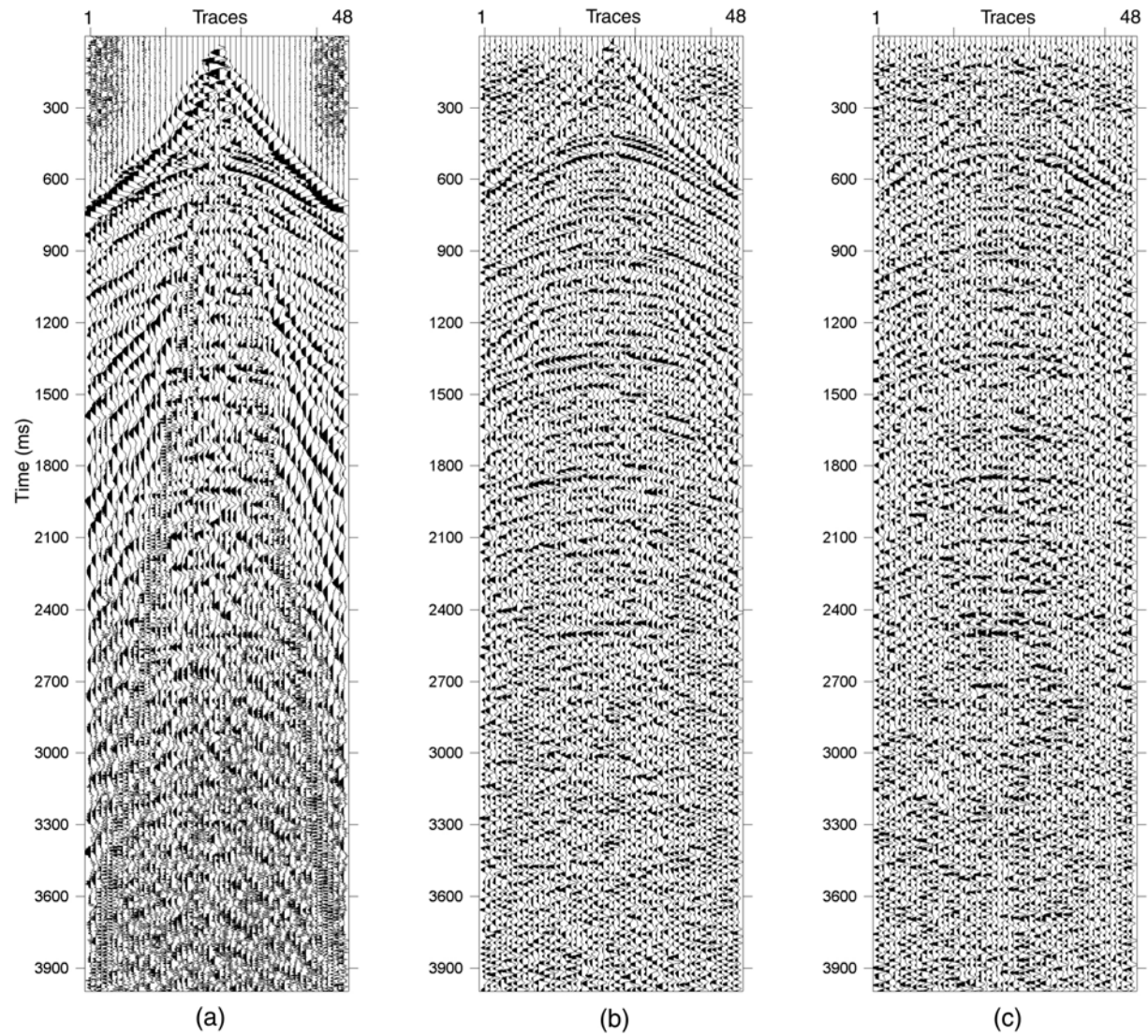

Figure 4